\documentclass[12pt,showpacs,preprintnumbers,aps]{revtex4}

\usepackage{amssymb}

\usepackage{epsf}
\usepackage{bm}

\begin{document}

\title{
About possibility 
to measure an electric dipole moment (EDM) of
nuclei
in the range $10^{-27} \div 10^{-32}$ $e \cdot cm$ 
in experiments for search of time-reversal violating generation of magnetic
and electric fields.
}
\author{ V. G. Baryshevsky}
\affiliation{Research Institute for Nuclear Problems, Belarusian State University, 
11 Bobryiskaya str., 220050, Minsk, Republic of Belarus,\\ E-mail: bar@inp.minsk.by}

\date{\today}

\begin{abstract}
The possibility 
to measure an electric dipole moment (EDM) of
nuclei in the range $10^{-27} \div 10^{-32}$ $e \cdot cm$ 
in experiments for search of time-reversal violating generation of magnetic
and electric fields is discussed.
\end{abstract}

\maketitle

Violation of the time reversal invariance of nature laws provides for elementary
particles (nuclei, atoms) the possibility to possess an additional quantum
characteristic - the electric dipole moment (EDM) $\overrightarrow{d}$,
which could exist along with other characteristics such as electric charge,
magnetic dipole moment, electric and magnetic polarisabilities. Plenty of
experiments set the limits for EDM of different particles, atoms and nuclei 
\cite{1,Cs,2,3,5,6}.
The attained experimental level gives, for example, for an electron EDM
$d_e < 1.6 \cdot 10^{-27}$ $e \cdot cm$ \cite{3}. The similar evaluations are
provided for nuclei. 
Meanwhile,
combine with calculations the experimental limits for $d$ can be
interpreted in terms of fundamental parity and time-invariance-violating
(P-,T-odd) parameters. These limits tightly constrain competing
theories of CP-violation.

So, it is very important to consider new possibilities for measuring constants
describing T and CP-odd interactions \cite{1,7,new_lanl,Mats}.
Very sensitive solid state based electron EDM experiments
are preparing now \cite{1}. They will provide for electron EDM measurement the sesitivity of 
about $10^{-32}$ $e \cdot cm$ or even better ($10^{-35}$ $e \cdot cm$) \cite{1}.

The idea of the experiment is based on the measurement of a magnetic field, which appears 
since the electron spin (and, therefore, the magnetic moment) is oriented along an external electric field $E$
due to interaction of electron EDM with the field $E$ \cite{5,6}.

According to \cite{1} measurement of electron EDM at the sensitivity rate of about
$10^{-32}$ $e \cdot cm$ requires substance cooling up to $T \sim 10^{-2}$ K, while for
$10^{-35}$ $e \cdot cm$ the temperature $T \sim 10 \mu K$ is necessary.

But it was shown in \cite{new_lanl} that for temperature values $T \approx 10^{-1} \div 10^{-2}$ 
the magnetic susceptibility
of matter $\chi$ becomes compartible with 1 and higher. 
The energy of interaction of two electron
magnetic dipoles for neighbour atoms occurs of order of $k_B T$ and greater ($k_B$ is Boltzmann's 
constant). 
Thus, in this case, the collective effects, well-known in the theory of phase transitions
in magnetism, should be taken into account while considering magnetization by an electric field.
Particularly, spontaneous magnetization of a system can appear. 
Emergent great magnetic field is caused by fluctuations of a magnetic field and 
weak residual magnetic fields rather than external electric fields. 
In such conditions
an electric field would not affect distinctly on the
value of the measured magnetic field. 
Therefore, measurement of electron EDM in such conditions
becomes difficult.

In the present paper it is shown that the above described problems make considerably 
attractive carrying out the experiment to search the EDM of nuclei 
by measurement of the magnetic
(electric) field
that appears at polarization of nuclear spins due to interaction of nuclear EDM with an external
electric (magnetic) field. 
The magnetic moment of a nucleus is small. 
Therefore, considering the degree of polarization 
to be the same for electrons and nuclei we obtain the value of this magnetic field for nuclei of
two or three orders lower then for electrons. But the phase transition of nuclear spins in 
an ordered state appears at temperatures $10^4 \div 10^5$ times lower comparing with electrons.
According to \cite{Abragam} the nuclear dipole ordering appears at $T_N \sim 10^{-6}$ K.
Let us consider the parameter $\varkappa = \frac{d_N E}{kT}$, which describes the degree of nuclei spin polarization
due to interaction of the nuclear EDM $d_N$ with an electric field $E$ in the temperature range above the temperature
of phase transition.
The temperature $T_N$ is $10^{4} \div 10^{5}$ times lower 
than the similar temperature for electrons $T_e$ ($T_e \sim 10^{-1} \div 10^{-2}$ K).
Hence, for nuclei $\varkappa$ is $10^4 \div 10^5$ times greater than for electrons.

As a result, the magnetic field produced by nuclei cooled to the temperature $T_N$
is even stronger than that produced by electrons cooled to the temperature $T_e$.
As a consequence, the nuclear EDM can be measured with sensitivity about $10^{-32}$ $e \cdot cm$.

\section{Nuclei in an electric field at low temperatures}

Thus,let us consider a substance, placed into an electric field. 

Interaction $W_E$
of an EDM $\overrightarrow{d}_{N}$ of a nucleus with an electric field $\overrightarrow{E}$
is similar to interaction of magnetic moment of a nucleus with a magnetic field and can be expressed as:
\begin{equation}
W_E=-\overrightarrow{d}_{N}\overrightarrow{E}
\label{1}
\end{equation}
where $\overrightarrow{d}_{N}=d_{N} \frac{\overrightarrow{J}}{J}$, $\overrightarrow{J}$ is the nucleus spin.

If the nucleus environment in the substance does not possess cubic symmetry, then 
the energy of interaction of nuclear quadrupole moment with a nonuniform electric field field
produced nucleus environment should be added to (\ref{1}).
If an external magnetic field also presents, then the energy of interaction of nuclear magnetic moment 
with this field should be added to (\ref{1}), too.

But here
we will omit these contributions, assuming that there is no an external magnetic field and 
an elementary cell of the substance is cubic.

The interaction of nucleus with the field $\overrightarrow{E}$ (\ref{1}) makes spins of nuclei
at low temperature polarized similar to the polarization  (magnetization) of nuclei
by a magnetic field due to the interaction $W_B$ of a nucleus magnetic moment $\mu_{N}$ 
with a magnetic field $\overrightarrow{B}$:
\begin{equation}
W_B=-\overrightarrow{\mu}_{N}\overrightarrow{B}
\label{2}
\end{equation}

Spins of nuclei polarized by an electic field induce the magnetic field
$\overrightarrow{B}_{E}$ and change in the magnetic flux $\Phi$ at the
surface of a flat sheet of material (Fig.1):
\begin{eqnarray}
\Delta \Phi_N=4 \pi \chi_{N} A \frac{d_N}{{\mu}_{N}} E^{*} \\
B_E=\frac{\Delta \Phi_N}{A}=4 \pi \chi_{N} \frac{ d_N }{{\mu}_{N}}E^{*},
\label{3}
\end{eqnarray}
where $A$ is the sample area, $\chi_{N}$ is the magnetic susceptibility of nuclei subsystem.
\begin{figure}[htbp]
\epsfxsize = 14.5 cm \centerline{\epsfbox{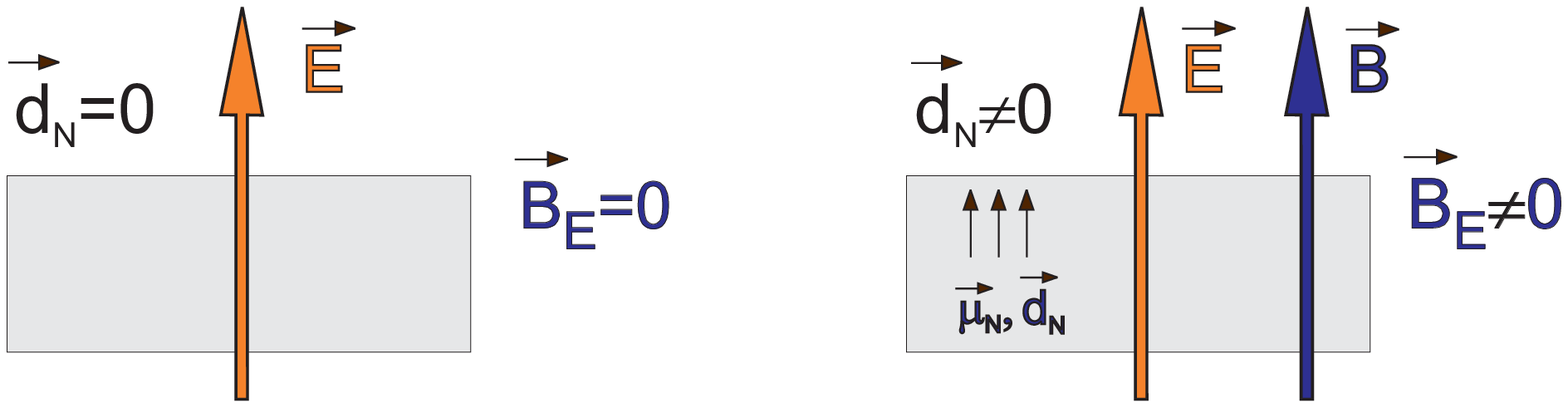}}
\caption{}
\end{figure}
For the temperature range above the temperature of nuclear magnetic ordering the susceptibility
$\chi_N$ is described by Langevin's formula as follows:
\begin{equation}
\chi_N \approx \frac{\rho \mu_{N}^2}{3 k_B T},
\label{5}
\end{equation}
$\rho$ is the number of nuclear spins in cm$^3$, $k_B$ is Boltzmann's constant, 
$\mu_{N}=g_N \sqrt{J(J+1)} \mu_{NB}$, $\mu_{NB}$ is the nucleus Bohr magneton, $g_N$
is the gyromagnetic ratio, $E^{*}$ is the effective electric field at the location of
the nuclear spin. Reasons providing interaction of an electric dipole moment with
an electric field in spite of electrostatic shielding are specified in \cite{1,9}.

Let us compare (\ref{3}) for $\Delta \Phi_N$ with the similar expression in \cite{1}
for the magnetic flux $\Delta \Phi_e$ produced by electrons of atoms:
\begin{equation}
\Delta \Phi_e=4 \pi \chi_{e} A \frac{d_e}{{\mu}_{a}} E^{*},
\label{7}
\end{equation}
where $\chi_{e}$ is the magnetic susceptibility of a material caused by paramagnetic atoms, 
$\mu_a$ is the magnetic moment of the electron shell of the atom.

As it has been already mentioned, the magnitude of the susceptibility $\chi$ grows with
temperature $T$ tending to the temperature of magnetic ordering and appears equal
to 1 at some temperature value $T_1$. At this temperature 
(note that for nuclei $T_{1N} \sim 10^{-6}$ K, while for atom spins 
$T_{1} \sim 10^{-1} \div 10^{-2}$ K) the magnetic flux produced by nuclei is
\begin{eqnarray}
& & \Delta \Phi_N (T_{1N})=4 \pi A \frac{d_N {E^{*}}_N}{{\mu}_{N}} ,\\
& & T_{1N} \sim 10^{-6}~K, \nonumber
\label{8}
\end{eqnarray}
the magnetic flux produced by atoms
\begin{eqnarray}
& & \Delta \Phi_e (T_{1e})=4 \pi A \frac{d_e {E^{*}}_a}{{\mu}_{a}} ,\\
& & T_{1e} \sim 10^{-1} \div 10^{-2} ~K. \nonumber
\label{9}
\end{eqnarray}

Let us consider temperature range close to the temperature of nuclear magnetic ordering.
Suppose nuclear EDM is of the same order as EDM of an electron (atom) 
$d_N {E^{*}}_N \sim d_e {E^{*}}_a$. 

Then, from (\ref{8},\ref{9}) we obtain
the magnetic flux 
$\Delta \Phi_N (T_{1N})=\frac{{\mu}_{a}}{{\mu}_{N}}\Delta \Phi_e (T_{1e})$ i.e. the magnetic flux produced
by nuclei is $\frac{{\mu}_{a}}{{\mu}_{N}}$ times greater than the magnetic flux
produced by atom spins!

According to estimations \cite{1}  the sensitivity
for atom EDM measurement $\sim 10^{-30}~e \cdot cm$ in 10 days of averaging (and even
$10^{-32}~e \cdot cm$) can be obtained for the material temperature $T_a=10^{-2}$ K.

Cooling of nuclear system to the temperature $T_N \approx 10^{-5}$ K provides
the magnetic flux $\Delta \Phi_N=\frac{{\mu}_{N}}{{\mu}_{a}} \frac{T_a}{T_N} \Delta \Phi_e$.
Due to $\frac{\mu_a}{\mu_N} \frac{T_a}{T_N} \approx 1$ the sensitivity for measurement of nuclear EDM appears the same as for
electron (atom) EDM i.e. $d_N \approx 10^{-30} \div 10^{-32}~ e \cdot cm$.

Thus, the method for measurement of nuclear EDM by means of measurement of the induced
magnetic field gives hope to obtain
considerably substantial limit for the nuclear EDM.
Let us note that in ferroelectric materials there are very intense internal electric fields
$E*$ that are considerably more intense than external fields. This provides to advance to 
$d_N$ values less than $10^{-32}~e \cdot cm$.

\section{Nuclei in a magnetic field at low temperatures}

If an external magnetic field acts on a material,
the nuclei spins become polarized due to nucleus magnetization. Therefore, the nucleus electric dipole
moments appears polarized, too. This results in the induction of an electric field $\overrightarrow{E}_{B}$
(Fig.2)
(similar the  electron case \cite{1}):
\begin{equation}
E_B=4 \pi \rho d_N P(B),
\label{6}
\end{equation}
where $P$ represents the degree that the spins of nuclei are polarized in the sample.

There are methods providing $P(B) \sim 1$ for nuclei at low temperatures.
\begin{figure}[htbp]
\epsfxsize = 14.5 cm \centerline{\epsfbox{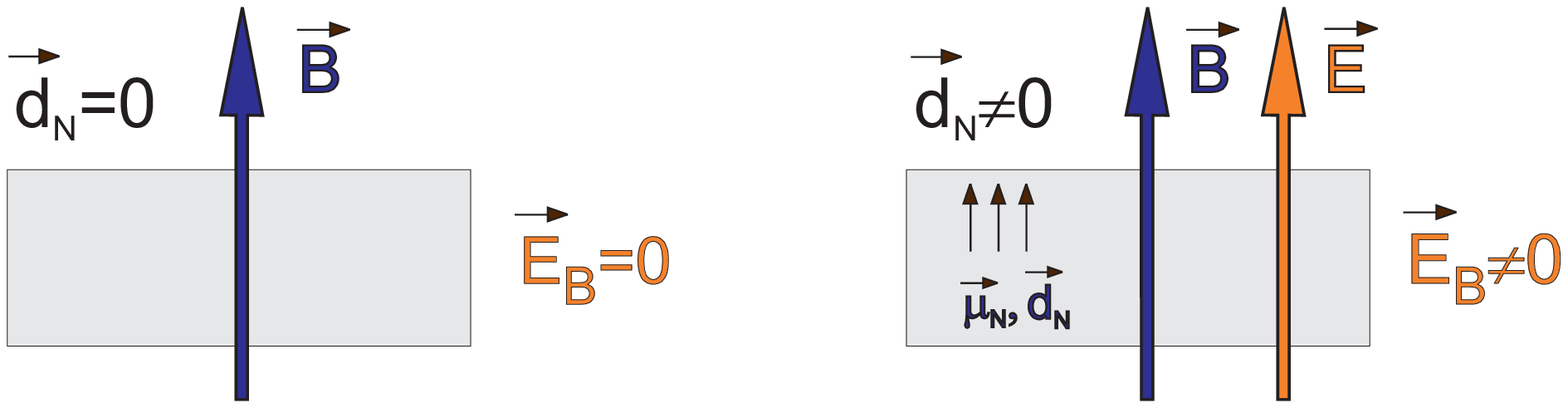}}
\caption{}
\end{figure}


According to the analysis \cite{1}
the technique for electric fields measurement provides a sensitivity
$\sim 10^{-30}~e \cdot cm$ in 10 days of operation.
The same sensitivity can be obtained for nuclear EDM due to high polarization degree
available for nuclear spins at low temperatures.

An electric field induced by a magnetic field in vacuum can be increased by the use of
electrostatic shielding or by selection of target shape (for example, a target having an edge).
This gives hope for further improvement of the possibile limits for the EDM measurement 
($d<10^{-30}$ $e \cdot cm$)
in such experiments.

\section{Time-reversal violating generation of static magnetic and electric fields}

It is important to pay attention 
that carrying out the proposed experiment to measure nuclear EDM
one should consider additional effect of 
time-reversal violating generation of static magnetic and electric fields caused by
the T-odd polarizability of atoms, molecules and other particles \cite{7,new_lanl,Mats}.
According to the idea of \cite{7}, 
an induced 
magnetic field
appears on the particle due to the action of a field $\vec E$ 
under conditions of violation of P- and T-invariance 
(and similar, an induced electric field 
appears on the particle due to the 
action of a field $\vec B$).
This new effect does not depend on temperature.
An effect magnitude is determined by 
a P-odd T-odd tensor polarisability $\beta_{ik}^{T}$
of a particle (atom, molecule, nucleus, neutron, electron and so on).
For an atom (molecule), $\beta_{ik}^{T}$ arises due to P- and T-odd 
interaction of electrons with a nucleus \cite{7,new_lanl,Mats}.

Let us place an atom (molecule) into an electric field $\vec {E}$.
The induced magnetic dipole moment $\vec{\mu} (\vec E)$
appears in this case \cite{7}:
\begin{equation}
{\mu_{i} (\vec E)}=\beta_{ik}^{T} {E}_k,
\end{equation}
The tensor $\beta_{ik}^{T}$ (like any tensor of rank two) can be expanded into scalar, 
simmetric and antisimmetric parts.

The antisymmetric part of the tensor $\beta_{ik}^{T}$ is proportional to
$e_{ikl} J_l$, where $e_{ikl}$ is the totally antisymmetric tensor of rank three.
The symmetric part of the tensor $\beta_{ik}^{T}$ is proportional to the tensor of 
quadrupolarization
$Q_{ik}=\frac{3}{2J(2J-1)}[J_i J_l+J_k J_l-\frac{2}{3} J(J+1)\delta_{ik}]$.
As a result
\begin{equation}
\beta_{ik}^{T}=\beta_{s}^{T} \delta_{ik} + \beta_{v}^{T} e_{ikl} J_l + \beta_{t}^{T} Q_{ik},
\end{equation}
where
$\beta_{s}^{T},\beta_{v}^{T}, \beta_{t}^{T}$ are the scalar, vector and tensor 
P-, T-odd polarizabilities of the particle, respectively.
For a substance with the nonpolarized spins
$Sp~\rho(J){\vec J}=0$ and $Sp~\rho (J) Q_{ik}=0$ (here $\rho (J)$ is the atom 
(molecule) spin density matrix).
As a result for such a substance, $\beta_{ik}^{T}$ appears to be a scalar
$\beta_{ik}^{T}=\delta_{ik} \beta_{s}^{T}$.

It follows from (\ref{8}) that in a substance placed into
electric field the magnetic field is induced \cite{7}:
\begin{equation}
\vec{B}^{ind}(E)=4 \pi \rho \beta _{ik}^{T} \vec{E}^{*}_k,
\label{9}
\end{equation}
where $\vec{E}^{*}_k$ is the local electric field acting on an atom in the substance.

Vice versa, if an atom (molecule, nucleus) is placed into a magnetic field, 
the induced electric dipole moment $\vec{d}(B)$
appears \cite{7}, 
\begin{equation}
d_{i}(B)=\chi_{ik}^T B_{k},
\label{d_i}
\end{equation} 
where
the tensor polarizability
$\chi_{ik}^T$
is  $\chi_{ik}^T=\beta_{ki}^T$.  
The above dipole moment $\vec{d}(B)$ leads to the induction of an electric field 
in the substance:
\begin{equation}
{E}_{i}^{ind}(B)=4 \pi \rho \beta _{ki}^{T} \vec{B}_{k}^{*} ,
\label{10}
\end{equation}
where $\vec{B^{*}}$ is the local magnetic field, 
acting on the considered particle in the substance.

Polarized electron spins give one more contribution to the magnetic field

Therefore, the magnetic field $\vec{B^{*}}$, which is going to be measured in the 
proposed experiment, should be written as:
\begin{equation}
\vec{B^{*}}  =  \vec{B}_{E}^N+\vec{B}^{ind}(E)+\vec{B}_{E}^A,
\label{12}
\end{equation}
where $\vec{B}_{E}^A$ is the magnetic field produced by atom spins.

Let us consider now the experiment to detect the electric dipole moment of the nucleus
by means of measurement of the electric field (see (\ref{5})). 
In this case we also should take into consideration
the effect of electric field induction by the magnetic field (\ref{d_i}) and the electric field ${E_B^A}$
produced by polarized EDM of atoms.

Thus, the electric field measured in the proposed experiment is as follows
(see (9),(10)):
\begin{equation}
\vec{E}_{B}=\vec{E}_B^N+\vec{E}^{ind}(B)+\vec{E}_B^A.
\label{13}
\end{equation}

So, measurement of $\vec{B}_{E}$ and $\vec{E}_{B}$ 
provides knowledge about
nuclear EDM, electron EDM
and $\beta _{s}^{T}$.
To distinguish these contributions one should consider the fact that 
studying $B_E$ and $E_B$ dependence on temperature allows one to
evaluate different contributions from \cite{new_lanl}.

It should be emphasized that atom EDM does not contribute to the discussed phenomena
if the substance is diamagnetic (atom spin is equal to zero or spin atoms are nonpolarized) \cite{new_lanl}. 

If the substance consists of several types of atoms (nuclei), then
their contribution to the induced field is expressed as a sum of contributions from different atoms:
\begin{equation}
\vec{B}_E= \sum\limits_{n} \vec{E}_n^{*}, 
\vec{E}_B=\sum\limits_{n} \vec{B}_{n}^{*},
\label{*}
\end{equation}

\section{About contribution of electromagnetic interaction to the electric dipole moment of nuclei}

Analysis of experimental limits obtained for EDM of atoms and nuclei was done in \cite{9,Kozlov,Dmitriev}.
The possible values for constants describing T-odd, P-odd interaction
were found there on the basis of calculation Shiff EDM of nuclei  \cite{Kozlov,Dmitriev}. The structure of an electric field inside
a nucleus was also analysed
and it was shown that it leads to an additional contribution to nuclear EDM, which was called LDM \cite{Kozlov}.
The mentioned contributions to EDM are finally caused by interference of strong, T-,P-odd and electrostatic 
interactions. 
Attention should drawn to the fact that there is one more contribution to nuclear EDM different from those, 
considered in \cite{Kozlov,Dmitriev}. This addition is caused by electromagnetic currents \cite{new_lanl}.

According to the above the induced magnetic moment
$\overrightarrow{\mu}_{E}$ of a particle appears due to action of a field 
$\overrightarrow{E}$ under conditions of violation of P- and T-invarince (and similar,
induced electric dipole moment $\overrightarrow{d}_B$ of a particle appears due to the action
of a magnetic field $\overrightarrow{B}$):
\begin{equation}
\mu^E_i=\beta^T_{ik}{E_k}, \\
\end{equation}
\begin{equation}
d^B_i=\beta^T_{ki}{B_k},  
\label{interaction}
\end{equation}
where $\beta^T_{ik}$ is the T-odd polarizability tensor 
of the particle (atom, nucleus, neutron, electron and so on).

As a result, the induced electric $\overrightarrow{E}_{ind}(\overrightarrow{r})$ and
magnetic $\overrightarrow{B}_{ind}(\overrightarrow{r})$ fields appear  in space.
The magnetic moment of a nucleus (multipoles of high orders) interacts with the field
$\overrightarrow{B}_{ind}(\overrightarrow{r})$. Particularly, interaction of the magnetic moment density
$\overrightarrow{\mu}(\overrightarrow{r})$ of a particle with the field
$\overrightarrow{B}_{ind}(\overrightarrow{r})$ can be expressed as:
\begin{equation}
W_{ind}=-\int \overrightarrow{\mu}(\overrightarrow{r})
\overrightarrow{B}_{ind}(\overrightarrow{r})
d^3 
\label{Wind}
\end{equation}

As the field $\overrightarrow{B}_{ind}(\overrightarrow{r})$ is proportional to the
electric field $\overrightarrow{E}$, then (\ref{Wind}) can be rewritten
\begin{equation}
W_{ind}={\chi_{_{TN}}}{\mu}_N \frac{\overrightarrow{J}}{J}\overrightarrow{E},
\label{Wind1}
\end{equation}
where constant $\chi_{_{TN}}$ is the T-odd susceptibility of a nucleus,
$\mu_N$ is the magnetic moment of the nucleus.

The susceptibility $\chi_{_{TN}}$ can be evaluated as \cite{7,new_lanl}:
\begin{equation}
\chi_{_{TN}} \sim \frac{\beta_{TN}}{a^3},
\label{chiTN}
\end{equation}
where $a$ is the typical radius of a nucleus.

As one can see (\ref{Wind1}) is similar 
to the interaction of an electric dipole moment 
$d_{\mu}=\mu_N \chi_{_{TN}}$ with an electric field.

One more contribution to the nuclear EDM appears by the following way.
The magnetic moment of a nucleus $\mu_N$ creates the magnetic field 
$B_N \sim \frac{8 \pi}{3} \frac{\rho}{A} \mu_N$ inside the nucleus, here $\frac{\rho}{A}$
is the nucleus density per one nucleon. According to
(\ref{interaction}) this field induces an additional contribution to the EDM
$d \sim \beta_T B_N$.

More rigorously the mentioned contributions to the EDM can be calculated by including 
electromagnetic interactions between nucleons
in Hamiltonian along with strong and weak interactions and by considering of radiation corrections to
the energy of interaction of a nucleus with an external field \cite{new_lanl}.

As a result the total energy of nucleus interaction with an
external electric field can be written as follows: 
\begin{equation}
W=-d \frac{\overrightarrow{J}}{J} \overrightarrow{E}-\chi_{T} \mu \frac{%
\overrightarrow{J}}{J} \overrightarrow{E}= -\frac{1}{J}(d+\chi_{T} \mu)%
\overrightarrow{J}\overrightarrow{E}=
-\frac{1}{J}(d+d_{\mu})\overrightarrow{J}\overrightarrow{E}
=-D_N \frac{\overrightarrow{J}%
\overrightarrow{E}}{J},  \label{W_tot}
\end{equation}
where $d$ is the contribution to EDM considered in \cite{Kozlov,Dmitriev},
$d_{\mu}$ is the contribution to EDM due to magnetic field induced inside the nucleus
(let us call it pseudo-dipole moment $d_{\mu}$), $D_N=d+d_{\mu}$ is nucleus EDM.

Now, let us estimate possible values of the susceptibility $\chi_{_{TN}}$ and
pseudo-dipole moment $d_{\mu}$ (PDM). 

An induced
magnetic field $B_{ind} \sim \frac{\mu_{ind}}{a^3}$ ($a$ is some
typical radius of the density distribution, for a nucleus it could be the radius of the nucleus).

Interaction of the magnetic moment $\mu$ of a particle with the field $%
B_{ind}$ can be estimated as $W_{ind} \approx \frac{\mu \mu_{ind}}{a^3}$. The
induced magnetic moment 
\begin{equation}
\mu_{ind} \sim \beta_T E \sim \frac{\left\langle d\right\rangle
\left\langle \mu \right\rangle}{\Delta} \eta_T E,
\end{equation}
where $\left\langle
d\right\rangle$ is the transition matrix element of the operator of electric dipole
moment, $\left\langle \mu \right\rangle$ is the transition matrix element of the
operator of magnetic dipole moment, $\Delta$ is the typical distance between
levels of opposite parity (for neutron it is about 1 GeV, for nuclei it is about hundreds keV $\div$ MeV), 
$\eta_T$ is the
coefficient of mixing of the opposite parity states by T-,P-odd interaction, 
$\eta_T \approx \frac{V_{TP}}{\Delta}$, $V_{TP}$ is the matrix element of
T-,P-odd interaction between states of opposite parity.

Therefore, 
\begin{eqnarray}
& & W_{ind}\sim \frac{\mu \mu _{ind}}{a^{3}}\sim \mu \frac{\left\langle
d\right\rangle \left\langle \mu \right\rangle }{a^{3}\Delta }\eta_{T}E =\mu
\chi_T E , \\
& & {\text{the susceptibility }} \chi_T \sim \frac{\beta_T}{a^3} \sim \frac{\left\langle d\right\rangle
\left\langle \mu \right\rangle }{a^{3}\Delta }\eta_{T},  \label{14} \\
& & d_{\mu}=\mu \chi_T \sim \chi_T {\lambda_c}({\text{cm}} \cdot e). 
\label{20}
\end{eqnarray}

From (\ref{20}) the following estimation for an electromagnetic contribution $d_{\mu N}$ 
to nucleus EDM can be odtained.

Considering a ''free'' nucleus (without electron shell) we can obtain from (\ref{20})
the following estimation for the addition $d_{\mu N}$ induced by magnetic
interactions 
\begin{equation}
d_{\mu N}=\mu _{N}\chi _{NT}\sim A \frac{V_{coul}}{\Delta }\frac{\lambda _{c}}{%
a}\eta _{NT}~\lambda _{c}~~{\text{cm}} \cdot e,
\end{equation}%
$a=a_0 A^{\frac{1}{3}}~ (a_0=1.2 \cdot 10^{-13}~cm)$ is the nucleus radius, $A$ is the number 
of nucleons in the nucleus, for this estimation it is suposed that $d \sim e a$, $\mu \sim \frac{e \hbar%
}{mc}=e \lambda_c$, $\lambda_c=\frac{\hbar}{mc}$ is the Compton wavelength of the nucleon, $%
V_{coul}=\frac{e^2}{a}$. 

Suppose
that typical difference between levels is about several MeV we obtain for heavy nuclei
\begin{equation}
d_{\mu N}\sim 10^{-15}\eta _{NT}~e \cdot cm,
\end{equation}%

This estimation is done in assumption that nucleons are unstructured. Considering also 
a T-odd polarizability $\beta_{N}$ of nucleons we can see that an external electric field
induces a nucleon magnetic moment $\mu_{N~ind}=\beta^T_{N}E$ and this magnetic moment creates
a magnetic field inside the nucleus
\begin{equation}
B_{N~ind} \sim \frac{8 \pi}{3} \rho \mu_{N~ind}=\frac{8 \pi}{3} \rho \beta^T_{N}E,
\end{equation}%
The interaction of the magnetic moment of the nucleus with this field contributes
to nucleus EDM as follows:
\begin{equation}
d_{N~\mu} \sim \frac{8 \pi}{3} \rho \mu_{N~ind}=\frac{8 \pi}{3} \rho \beta^T_{N}E,
\end{equation}%
where $\rho$ is the density of nucleons in the nucleus, for a nucleon 
$\beta^T_{N} \sim \frac{V_{coul}^N}{\Delta } \eta_T^N {\lambda _{c}}^3$.
Suppose that for a nucleon $a \sim \lambda _{c}$, $\Delta \sim 1 Gev$, 
$V_{coul}^N \sim \frac{e^2}{\lambda _{c}}$.

Therefore, the additional contribution to the dipole moment of a nucleus aroused from
the T-odd polarizability of nucleons is:
\begin{equation}
d_{N~\mu} \sim \frac{8 \pi}{3} \rho {\lambda _{c}}^3 \frac{V_{coul}^N}{\Delta} \eta_T^N
\sim 10 \lambda _{c} \rho {\lambda _{c}}^3 \frac{V_{coul}^N}{\Delta} \eta_T^N 
\approx 
10  \rho {\lambda _{c}}^3 10^{-2}\lambda _{c} \eta_T^N (cm \cdot e),
\end{equation}
for nuclei with $A>20$ the density $\rho=10^{38}$, i.e. 
$d_{N~\mu} \approx 10^{-18} \eta_T^N (cm \cdot e)$

The addition $d_{N~\mu}$ can be expressed by means of $d_{n~\mu}$ of nucleon:
\begin{equation}
d_{N~\mu} = \frac{8 \pi}{3} \rho {\lambda _{c}}^3 d_{n~\mu} \approx 10^{-2} d_{n~\mu},
\end{equation}

The same conclusion can be done from the following reasoning. The magnetic moment of a nucleus
$\mu_N$ creates inside the nucleus a magnetic field 
$B_{ind} \sim  \frac{8 \pi}{3} \frac{\rho}{A} \mu_N$, where $\frac{\rho}{A}$ is the nucleus density
per one nucleon. The field $B_{ind}$ induces an electric dipole moment of nucleon
\begin{equation}
d_{ind} = \beta^T_{N} B_{ind},
\end{equation}

Total induced EDM is $D_{ind}=Ad_{ind}=\frac{8 \pi}{3} \rho \beta^T_{N} \mu = d_{N~\mu}$

It should be mentioned that the spin structure of the nuclear interaction between nuclons is similar 
to the magnetic interaction. {Due to this reason
the spin of the nucleus acts on the other nucleons by dint of the average nuclear field, depending
on the spin orientation (let us call this field a pseudo-magnetic nuclear field, similar to the field 
appearing in a polarized nuclear target).} The action of this field 
(similar to the action of an ordinary magnetic field) induces the electric dipole moment of nucleons 
in the nucleus. This contribution is appreciably (two orders) greater than $d_{N~\mu}$ due to 
big value of the pseudo-magnetic nuclear field, but it is quite difficult to evaluate this contribution.

Let us consider deuteron EDM.
Deuteron EDM was calculated in \cite{14}. According to \cite{14} deuteron EDM
arouses due to mixing of stationary states of a system ''neutron+proton'' 
{(interacting one with each other by strong interaction)} by
the T-odd, P-odd interactions.

According to the above analysis there are some additions to the deuteron EDM caused 
by the electromagnetic interactions between proton and neutron and nucleon polarizabilities.

{
Considering only strong interactions we can find a wavefunction of P-state of a system of two nucleons 
using an approximation of zero radius of nuclear forces, because strong interactions are short-range \cite{14}. 
But electromagnetic interactions makes the interaction between n and p far-ranging.
Thus, taking electromagnetic interaction into account we must consider for P-state a correcton to the 
wavefunction, caused by this interaction.}

A matrix element of the transition current for the n-p system can be expressed in conventional form:
\begin{eqnarray}
\overrightarrow{j}_{NF}(\overrightarrow{r})=
\frac{i e \hbar}{2m}\left( \Psi _{F}\overrightarrow{\nabla }_{r}\Psi _{N}^{\ast }-\Psi
_{N}^{\ast }\overrightarrow{\nabla }_{r}\Psi _{F}\right) -\frac{e^{2}}{mc}%
\Psi _{N}(\overrightarrow{r})\overrightarrow{A}(\overrightarrow{r})\Psi
_{F}+\\
c~rot\left[ \Psi _{N}^{\ast }\left( \frac{1}{2}\left( \mu _{p}+\mu
_{n}\right) \left( \overrightarrow{\sigma }_{p}+\overrightarrow{\sigma }%
_{n}\right) +\frac{1}{2}\left( \mu _{n}-\mu _{p}\right) \left( 
\overrightarrow{\sigma }_{n}-\overrightarrow{\sigma }_{p}\right) \right)
\Psi _{F}\right]. \nonumber
\end{eqnarray}
the term proportional to 
$\left( \overrightarrow{\sigma }_{n}+\overrightarrow{\sigma }_{p}\right)$
describes transitions between triplet states, while the term including
$\left( \overrightarrow{\sigma }_{n}-\overrightarrow{\sigma }_{p}\right)$
corresponds to the transitions between triplet and singlet states.
Let us place deuteron into an external electric field. The Hamiltonian
of the n-p system is expressed as:
\begin{equation}
H = H_0+V_E+V_W,
\end{equation}
where $H_0$ is the  Hamiltonian of the np system considering both the strong and 
electromagnetic interactions between nucleons,
$V_E=-\overrightarrow{d}\overrightarrow{E}$, 
$\overrightarrow{d}=\frac{e}{2}\overrightarrow{r}$ is the operator of the electric 
dipole moment of the n-p system.
According to \cite{13} the energy of T-odd interaction
\begin{eqnarray}
& & V_{W}=-\frac{g\,g_{1}}{4\pi m_{p}}\overrightarrow{J}\overrightarrow{\nabla }%
\frac{e^{-m_{\pi }r}}{r}+\frac{3g\,g_{0}}{8\pi m_{p}}\left( \overrightarrow{%
\sigma }_{n}-\overrightarrow{\sigma }_{p}\right) \overrightarrow{\nabla }%
\frac{e^{-m_{\pi }r}}{r},\label{VW}\\
& & J=\frac{1}{2}\left( \overrightarrow{\sigma }_{n}+\overrightarrow{\sigma }_{p}\right) 
\nonumber
\end{eqnarray}

Calculating  the current $\overrightarrow{j}$ we should take the wavefuncton
$\Psi_{F(N)}$ considering contributions of the second order over 
$V_E+V_W$ interacton.

Considering only strong and weak interactions
for the deuteron EDM \cite{14}
one should take into account the first term proportional to 
$\left( \overrightarrow{\sigma }_{n}+\overrightarrow{\sigma }_{p}\right)$.
Consideration of
electromagnetic interactions leads to the conclusion that
the term proportional to
$\left( \overrightarrow{\sigma }_{n}-\overrightarrow{\sigma }_{p}\right)$
also contributes to the deuteron EDM. 

So, due to electromagnetic interaction the deuteron EDM is determined by both the constants 
$g\,g_{1}$ and $g\,g_{0}$.
Considering estimation (31-33) we obtain
for deuteron $d_{\mu D}\sim 10^{-16}\eta _{DT}$ cm$\cdot$e.

\end{document}